\documentclass[a4paper,11pt]{article}

\usepackage[utf8]{inputenc}
\usepackage{amsmath}
\usepackage{amssymb}
\usepackage{authblk}
\usepackage{booktabs}
\usepackage[cm]{fullpage}
\usepackage{graphicx}
\usepackage[colorlinks]{hyperref}
\usepackage{natbib}
\usepackage{pdflscape}
\usepackage{bm}
\usepackage[parfill]{parskip}

\setcitestyle{authoryear,open={(},close={)}}

\DeclareMathOperator{\Categorical}{Categorical}
\DeclareMathOperator{\Dirichlet}{Dirichlet}
\DeclareMathOperator{\Lognormal}{Lognormal}
\DeclareMathOperator{\N}{N}

\title{Investigating the efficiency of marginalising over discrete parameters
in Bayesian computations}

\author[1]{Wen Zhang}
\author[1,2]{Jeffrey Pullin}
\author[3]{Lyle Gurrin}
\author[1,2,*]{Damjan Vukcevic}

\affil[1]{School of Mathematics and Statistics,
          University of Melbourne, Australia}
\affil[2]{Melbourne Integrative Genomics,
          University of Melbourne, Australia}
\affil[3]{Melbourne School of Population and Global Health,
          University of Melbourne, Australia}
\affil[*]{Corresponding author: Damjan Vukcevic,
\href{mailto:damjan.vukcevic@unimelb.edu.au}{\nolinkurl{damjan.vukcevic@unimelb.edu.au}}}

\date{13 September 2022}

\begin{document}

\maketitle

\begin{abstract}
Bayesian analysis methods often use some form of iterative simulation such as
Monte Carlo computation.  Models that involve discrete variables can sometime
pose a challenge, either because the methods used do not support such variables
(e.g.\ Hamiltonian Monte Carlo) or because the presence of such variables can
slow down the computation.  A common workaround is to marginalise the discrete
variables out of the model. While it is reasonable to expect that such
marginalisation would also lead to more time-efficient computations, to our
knowledge this has not been demonstrated beyond a few specialised models.

We explored the impact of marginalisation on the computational efficiency for a
few simple statistical models.  Specifically, we considered two- and
three-component Gaussian mixture models, and also the Dawid--Skene model for
categorical ratings.  We explored each with two software implementations of
Markov chain Monte Carlo techniques: JAGS and Stan.  For JAGS, it was possible
to compare marginalised and non-marginalised versions of the same model with
the same samplers.

Our results show that marginalisation on its own does not necessarily boost
performance.  Nevertheless, the best performance was usually achieved with
Stan, which requires marginalisation. We conclude that there is no simple
answer to whether or not marginalisation is helpful. It is not necessarily the
case that, when turned `on', this technique can be assured to provide
computational benefit independent of other factors, nor is it likely to be the
model component that has the largest impact on computational efficiency.
\end{abstract}


\clearpage

\tableofcontents


\section{Introduction}

In a Bayesian framework, performing inference often involves the use of Markov
chain Monte Carlo (MCMC) methods. Recently, MCMC methods based on Hamiltonian
Monte Carlo (HMC), such as the `No U-Turn Sampler' (NUTS) algorithm
\citep{JMLR:v15:hoffman14a} implemented in the probabilistic programming
language Stan \citep{carpenter2017}, have become popular, replacing methods
based on Gibbs sampling in many domains. HMC-based methods require gradients of
the likelihood function to be computed with respect to all parameters in the
model. However, these gradients cannot be computed for discrete parameters,
potentially limiting the use of HMC-based methods. One common approach to
overcome this drawback is to \emph{marginalise} the discrete parameters out of
the model. HMC-based methods can then used to draw samples from the joint
posterior distribution of the marginalised model. Today, marginalisation is the
only way to fit models with discrete parameters in Stan. The widespread use of
both Stan and models with discrete parameters has therefore made
marginalisation a widely used technique in applied Bayesian modelling.

Recently, some of the authors of the current manuscript presented the R package
\texttt{rater}, which implements Bayesian versions of several statistical
models that allow analysis of repeated categorical rating data
\citep{pullin_statistical_2021}. The models are based on, and include, the
Dawid--Skene (DS) model \citep{dawid1979}, and as such they include a discrete
latent class parameter for each item. The \texttt{rater} package uses Stan to
perform inference, and marginalised versions of the models are employed. The
manuscript describing \texttt{rater} discusses in some detail the techniques of
marginalisation and conditioning in an effort to explain the mathematical basis
of the technique. In addition, other authors such as \citet{joseph_2020} have
sought to explain how to marginalise out discrete parameters in specific
models.

While marginalisation requires substantial mathematical effort, folk wisdom in
the Stan community suggests that fitting models with marginalisation is more
efficient than using Gibbs sampling. Indeed, the \emph{Stan User's Guide}
asserts that marginalisation can allow ``more efficient sampling on an
iteration-by-iteration basis'' \citep{stanguide}. To date, however, there has
been little empirical evidence for the idea that marginalisation can improve
computational efficiency. Recently, \citet{yackulic_need_2020} considered
marginalising out latent states in a variety of Bayesian population models such
as the Cormack--Jolly--Seber (CJS) model. In particular, they demonstrated that
marginalisation can improve computational performance by orders of magnitude in
a CJS model. In their comparison, the fastest model---implemented using
Stan---was more than one thousand times more efficient than the slowest
non-marginalised model. These results highlight the potential of
marginalisation, and Stan, to greatly increase the efficiency of Bayesian
inference for models with discrete parameters. It is not clear however, whether
the results presented by \cite{yackulic_need_2020} generalise to other models
that contain discrete parameters.

Whether or not marginalisation makes computation more efficient is of both
practical and theoretical interest. Practically, the  efficiency of
marginalisation would guide both whether marginalisation should be considered
as a possible way to speed up generic Bayesian computations, and decisions
about which software (i.e.\ JAGS vs Stan) and options within these software to
use. Theoretically, it is of interest to determine whether the claimed
improvements in efficiency suggested by the Rao--Blackwell Theorem actually
hold in practice. Finally, this work could guide whether future probabilistic
programming languages (such as SilcStan \citep{silcstan}) should implement
automatic marginalisation for some, or even all, models.

In this paper, we explore the impact of marginalisation on computational
efficiency for other models that contain discrete parameters: two- and
three-component Gaussian mixture models and the Dawid--Skene model for
categorical ratings. We implement various marginalised and non-marginalised
version of the models in JAGS and marginalised versions of the models in Stan.
We then compare the computational efficiency of the different inference
approaches across various simulation scenarios.


\section{Methods}

\subsection{Models and marginalisation}
\label{model}

Here we describe the Gaussian mixture models and the Dawid–Skene model that we
used for the generation and analysis of simulated data, including the prior
distributions we chose for Bayesian inference. We include two versions of the
likelihood functions: (i)~`full' versions, that include the discrete variables;
and (ii)~marginalised versions, where the discrete variables are marginalised
out.

\subsubsection{Gaussian mixture models}

A Gaussian mixture model assumes that each observation is generated from one of
a finite set of Gaussian distributions, referred to as the \emph{mixture
components}. The model uses a discrete latent variable as a category label to
indicate the mixture component from which each observation was generated.

Suppose that we have $n$ continuously valued observations, $x_1, \dots, x_n$.
We assume that each observation $x_i$ belongs to one of $K$ mixture components.
Associated with each $x_i$ is a discrete latent variable $z_i \in \{1, \dots,
K\}$ that indicates the mixture component from which $x_i$ was drawn. We assume
that $z_1, \dots, z_n$ are independent and identically distributed as follows.
For $i = 1, \dots, n$,
\begin{equation*}
z_i \sim \Categorical(\pi_1, \dots, \pi_K),
\end{equation*}
where $\pi_k \in [0,1]$ for $k = 1, \dots, K$ and $\sum_{k=1}^{K} \pi_k = 1$.
The parameter $\pi_k$ specifies the proportion of the population of
continuously valued observations (the ${x}_{i}$'s) belonging to the $k$th
mixture component.

We assume that $x_1, \dots, x_n$ conditional on $z_i, \dots, z_n$ are
independent and identically distributed as follows. For $i = 1, \dots, n$ and
$k = 1, \dots, K$,
\begin{equation*}
x_i \mid z_i=k \sim \N(\mu_k, \sigma^2).
\end{equation*}
The variance, $\sigma^2$, is assumed to be the same for all mixture components.
For $k \in \{1, \dots, K\}$, $\mu_k$ and $\sigma^2$ are unknown parameters.

For convenience, we combine the above quantities into vector formats: let
$\bm{x} = (x_1, \dots, x_n)$; $\bm{z} = (z_1, \dots, z_n)$; $\bm{\pi} = (\pi_1,
\dots, \pi_K)$ and $\bm{\mu} = (\mu_1, \dots, \mu_K)$.

The full likelihood function (that includes $z$) for this model is
\begin{equation*}
\Pr(\bm{x}, \bm{z} \mid \bm{\pi}, \bm{\mu}, \sigma^2)
  = \prod_{i=1}^{n} f(x_i \mid \mu_{z_i}, \sigma^2) \cdot \pi_{z_i},
\end{equation*}
where $f(x_i \mid \mu_{z_i}, \sigma^2)$ is the probability density function of
a Gaussian distribution with mean $\mu_{z_i}$ and variance $\sigma^2$.

The marginalised likelihood function, where we marginalise over $\bm{z}$, is
\begin{align}
\Pr(\bm{x} \mid \bm{\pi}, \bm{\mu}, \sigma^2)
 &= \sum_{\bm{z}} \Pr(\bm{x}, \bm{z} \mid \bm{\pi}, \bm{\mu}, \sigma^2)  \notag \\
 &= \sum_{{z}_{1} = 1}^{K} \sum_{{z}_{2} = 1}^{K} \dots \sum_{{z}_{n} = 1}^{K}
    \Pr(\bm{x}, \bm{z} \mid \bm{\pi}, \bm{\mu}, \sigma^2)  \notag \\
 &= \sum_{{z}_{1} = 1}^{K} \sum_{{z}_{2} = 1}^{K} \dots \sum_{{z}_{n} = 1}^{K}
    \prod_{i=1}^{n} f(x_i \mid \mu_{{z}_{i}}, \sigma^2) \cdot \pi_{{z}_{i}}  \notag \\
 &= \left(\sum_{{z}_{1} = 1}^{K} f(x_1 \mid \mu_{{z}_{1}}, \sigma^2) \cdot \pi_{{z}_{1}}\right)
    \left(\sum_{{z}_{2} = 1}^{K} f(x_2 \mid \mu_{{z}_{2}}, \sigma^2) \cdot \pi_{{z}_{2}}\right)
    \dots
    \left(\sum_{{z}_{n} = 1}^{K} f(x_n \mid \mu_{{z}_{n}}, \sigma^2) \cdot \pi_{{z}_{n}}\right)  \notag \\
 &= \prod_{i=1}^{n} \sum_{{z}_{i} = 1}^{K} f(x_i \mid \mu_{{z}_{i}}, \sigma^2) \cdot \pi_{{z}_{i}}.
\label{eq:marg-likelihood}
\end{align}
We place the following prior distributions on the parameters:
\begin{align*}
\bm{\pi} &\sim \Dirichlet(\bm{\alpha}),  \\
\sigma   &\sim \Lognormal(0, 1),  \\
\mu_1    &\sim \N(0, 10^2),  \\
\mu_k    &\sim \N(0, 10^2, \mu_{k-1}), \quad \text{for $k = 2, \dots, K$,}
\end{align*}
where $\bm{\alpha}$ is a vector of $1$'s with length $K$, and $\N(0, 10^2,
\mu_{k-1})$ is the truncated normal distribution where we use the truncation
$\mu_k > \mu_{k-1}$ to avoid label switching.  We chose $10^2$ for the variance
of the prior distribution after trying a range of values and selecting one that
improved convergence.

\subsubsection{Dawid--Skene model}

The Dawid--Skene model for categorical ratings is applicable to scenarios where
a set of items is classified into categories by one or more raters.
Specifically, we have $I$ items and $J$ raters. We assume that each item
belongs to one of $K$ categories. Associated with each item is a (discrete)
latent parameter $z_i \in \{1, \dots, K\}$ that indicates the true category for
item $i$. Associated with each item--rater pair is a rating $y_{i,j}$ for item
$i$ given by rater $j$\footnote{For simplicity, we assume that each item is
rated exactly once by each rater. The model naturally generalises to arbitrary
numbers of ratings and more complex experimental designs, see
\citet{pullin_statistical_2021}.}. Let $\bm{y}$ be the vector of all ratings.
We assume that $z_i, \dots, z_I$ are independent and identically distributed as
follows. For $i = 1, \dots, I$,
\begin{equation*}
z_i \sim \Categorical(\pi_1, \dots, \pi_K),
\end{equation*}
where $\pi_k \in [0,1]$ for $k = 1, \dots, K$ and $\sum_{k=1}^{K} \pi_k = 1$.
One can consider $\pi_k$ as the prevalence of category $k$ in the population
from which the items are sampled. We let $\bm{\pi} = (\pi_1, \dots, \pi_K)$ and
$\bm{z} = (z_1, \dots, z_I)$.

For all possible pairs of $i,j$ where $i \in \{1, \dots, I\}$, $j \in \{1,
\dots, J\}$, we assume that $y_{i,j}$ conditional on $z_i$ are independent and
identically distributed as follows
\begin{equation*}
y_{i,j} \mid z_i
  \sim \Categorical(\theta_{j,z_{i},1}, \dots, \theta_{j,z_{i},K})
\end{equation*}
where $\theta_{j,z_{i},k} \in [0,1]$ for $k = 1, \dots, K$ and $\sum_{k=1}^{K}
\theta_{j,z_{i},k} = 1$. $\theta_{j,z_{i},k}$ is the probability that rater $j$
rates an item of true category $z_i$ as being in category $k$. We let
$\bm{\theta}$ be the vector of all possible $\theta_{j, z_{i}, k}$'s.

For this model, the full likelihood function that includes $z$ is
\begin{equation*}
\Pr(\bm{y}, \bm{z} \mid \bm{\theta}, \bm{\pi})
  = \prod_{i=1}^{I} \left(\pi_{z_i} \cdot
                          \prod_{j=1}^{J} \theta_{j,z_i,y_{i,j}}\right).
\end{equation*}
The likelihood function where we marginalise over $z$ is
\begin{equation*}
\Pr(\bm{y} \mid \bm{\theta}, \bm{\pi})
  = \prod_{i=1}^{I}
    \left(\sum_{k=1}^{K}
          \left(\pi_{k}\cdot \prod_{j=1}^{J}\theta_{j,k,y_{i,j}}\right)\right).
\end{equation*}

We place weakly informative prior probability distributions on the parameters:
\begin{align*}
\bm{\pi}          &\sim \Dirichlet(\bm{\alpha}),  \\
\bm{\theta_{j,k}} &\sim \Dirichlet(\bm{\beta_k}),
\end{align*}
where $\bm{\alpha}$ is a vector of length $K$ with positive elements,
$\bm{\theta_{j,k}}$ is the vector $(\theta_{j,k,1}, \dots, \theta_{j,k,K})$,
$\beta$ is a $K \times K$ matrix with positive elements, and $\bm{\beta_k}$ is
the $k$th row of this matrix. Specifically, $\beta$ contains the elements:
\begin{equation*}
\beta_{k,k'} = \left\{
  \begin{array}{ll}
          Np           & \quad \textrm{if } k=k' \\
    \frac{N(1-p)}{K-1} & \quad \textrm{otherwise}
  \end{array}
\right.
\forall k,k' \in {1,\dots,K}.
\end{equation*}
For the hyper-parameters $\bm{\alpha}$, $N$ and $p$, we used the default values
from the R package \texttt{rater} \citep{pullin_statistical_2021}:
$\bm{\alpha}$ is a vector of $3$'s, $N = 8$ and $p = 0.6$. See
\citet{pullin_statistical_2021} for a description of how to interpret these
parameters.

\subsection{Data simulations}

Here we describe how we simulated each test dataset.  For each model and
experimental scenario, we simulated 5 replicate datasets.

\subsubsection{Two-component Gaussian mixture model}

We simulated observations from the following probability density function:
\begin{equation*}
g(x_i) = \pi_1 \cdot f(x_i \mid \mu_1, 2^2) +
         \pi_2 \cdot f(x_i \mid \mu_2, 2^2),
\end{equation*}
where $f(x_i \mid \mu, \sigma^2)$ is the probability density function of a
Gaussian distribution with mean $\mu$, variance $\sigma^2$ and mixture
proportions $(\pi_1, \pi_2)$. We define the distance between components by
$|\mu_2 - \mu_1|$, and also $\bm{\mu} = (\mu_1, \mu_2)$ and $\bm{\pi} = (\pi_1,
\pi_2)$.

For this model, we were interested in how the mixture proportions and distance
between components modified the impact of marginalisation. We simulated four
datasets, each of size 200. We set the values of $\mu_1$ and $\mu_2$ as in the
second column of \autoref{tab:two-component_model}.

\begin{table}
\caption{Data simulation parameters for the two-component Gaussian mixture
         model.}
\label{tab:two-component_model}
\smallskip
\begin{center}
\begin{tabular}{lccc}
\toprule
 Dataset & $\bm{\mu}$ & $\bm{\pi}$ & $|\mu_2 - \mu_1|$ \\
\midrule
 Dataset 1 & $(-5,5)$     & $(0.5,0.5)$ & $10$ \\
 Dataset 2 & $(-5,5)$     & $(0.9,0.1)$ & $10$ \\
 Dataset 3 & $(-2.5,2.5)$ & $(0.5,0.5)$ &  $5$ \\
 Dataset 4 & $(-2.5,2.5)$ & $(0.7,0.3)$ &  $5$ \\
\bottomrule
\end{tabular}
\end{center}
\end{table}

We paired the two options for the distances between components with three
options for the mixture proportions: one with equal proportions, one with
moderately imbalanced proportions and one with very imbalanced proportions.
See the third and fourth columns of \autoref{tab:two-component_model}.
The mixture proportion for Dataset $2$ was set to be highly imbalanced. This
allows us to investigate how discrete sampling for small probabilities impacts
marginalisation.
Plots of the density function for each dataset can be found in the first rows
of \autoref{fig:two-component-result} and
\autoref{fig:two-component-log-result}.

\subsubsection{Three-component Gaussian mixture model}

Similar to the above, we simulated observations from the following the
probability density distribution:
\begin{equation*}
g(x_i) = \pi_1 \cdot f(x_i \mid \mu_1, 2^2) +
         \pi_2 \cdot f(x_i \mid \mu_2, 2^2) +
         \pi_3 \cdot f(x_i \mid \mu_3, 2^2).
\end{equation*}
We assume $\mu_1 < \mu_2 < \mu_3$ and define the maximal distance between
components by $|\mu_3 - \mu_1|$. We consider two scenarios: (i)~where the
components are equidistant, $\frac{\mu_3 - \mu_2}{\mu_2 - \mu_1} = 1$; and
(ii)~where the components are non-equidistant, specifically $\frac{\mu_3 -
\mu_2}{\mu_2 - \mu_1} = \frac{2}{5}$.

For the three-component Gaussian mixture model, we index the simulations by
both the maximal distance between components and whether the three components
are equidistant or not.

We simulated eight datasets, each of size 200.  We set the value of $\bm{\mu}$
as in the second column of \autoref{tab:three-component_model}.  The
combinations of different mixture proportions, maximal distances between
components and whether the components are equidistant or not gives us a $2
\times 2 \times 2$ balanced design, as shown in
\autoref{tab:three-component_model}.

\begin{table}
\caption{Data simulation parameters for the three-component Gaussian mixture
         model.}
\label{tab:three-component_model}
\smallskip
\begin{center}
\begin{tabular}{lcccc}
\toprule
 Dataset & $\bm{\mu}$         & $\bm{\pi}$         & $|\mu_3 - \mu_1|$ &
 Equidistant components \\
\midrule
 Dataset 1 & $(-10.5,0,10.5)$ & $(0.33,0.33,0.33)$ & $21$ & Yes \\
 Dataset 2 & $(-10.5,0,10.5)$ & $(0.5,0.3,0.2)$    & $21$ & Yes \\
 Dataset 3 & $(-7,0,7)$       & $(0.33,0.33,0.33)$ & $14$ & Yes \\
 Dataset 4 & $(-7,0,7)$       & $(0.5,0.3,0.2)$    & $14$ & Yes \\
 Dataset 5 & $(-6,0,15)$      & $(0.5,0.3,0.2)$    & $21$ & No  \\
 Dataset 6 & $(-6,0,15)$      & $(0.33,0.33,0.33)$ & $21$ & No  \\
 Dataset 7 & $(-4,0,10)$      & $(0.33,0.33,0.33)$ & $14$ & No  \\
 Dataset 8 & $(-4,0,10)$      & $(0.5,0.3,0.2)$    & $14$ & No  \\
\bottomrule
\end{tabular}
\end{center}
\end{table}

Plots of the density function for each dataset can be found in the first rows
of \autoref{fig:three-component-result} and
\autoref{fig:three-component-log-result}.

\subsubsection{Dawid--Skene model}

We simulated a dataset with $5$ raters and $100$ items, where each items
belongs to one of the $5$ categories ($J = 5$, $I = 100$, $K = 5$). We set the
prevalence of each category to be uniform, i.e.\ $\bm{\pi} = ({\frac{1}{5},
\frac{1}{5}, \frac{1}{5}, \frac{1}{5}, \frac{1}{5}})$.

For $j \in \{1, \dots, 5\}$ and $k \in \{1, \dots,5\}$, we let $\theta_{j,k,k}
= 0.7$ and $\theta_{j,k,l} = 0.075$ for all $l \neq k$. This means all raters
classify the items correctly 70\% of the time, and otherwise make errors
uniformly amongst the other categories.

\subsection{Software implementations}

We explored the computational efficiency for each model--data pair using both
JAGS and Stan. Specifically, we used the main R interfaces to each one:
\texttt{rstan} \citep{rstan} and \texttt{rjags} \citep{rjags}. All of our
models were coded `by hand'. We used the R package \texttt{posterior}
\citep{posterior} for summarising output from both JAGS and Stan.

JAGS supports direct sampling of discrete parameters. In contrast, Stan cannot
be used when the posterior distribution explicitly contains discrete
parameters, so marginalisation is required. Using JAGS, we directly compared
marginalised and full versions of the same model (which we refer to as
\emph{jags-full} and \emph{jags-marg}), and also compared the performance of
the JAGS models with the (necessarily marginalised) Stan model (which we refer
to as \emph{stan}). All implemented likelihood functions are described
mathematically in \autoref{model}.

Source code for reproducing all of the results in this manuscript is available
on GitHub\footnote{\url{https://github.com/katezhangwen/Efficiency-of-marginalising-over-discrete-latent-parameters}}.

\subsubsection{Gaussian mixture models with JAGS}

For the two- and three-component Gaussian mixture models, we have two versions
of each of the full and marginalised models in JAGS, as follows.

\paragraph{Marginalised models:}
\begin{enumerate}
\item The \emph{jags-marg-inbuilt} model is marginalised with the
    \texttt{dnormmix} distribution from the \texttt{mix} module.
\item The \emph{jags-marg} model is marginalised manually as in
    \autoref{eq:marg-likelihood}.
\end{enumerate}

\paragraph{Full models:}
\begin{enumerate}
\item The \emph{jags-full} model is unmarginalised and allowed to use any
    samplers from the \texttt{base}, \texttt{bugs} and \texttt{mix} modules in
    JAGS.
\item The \emph{jags-full-restricted} model is unmarginalised but only allowed
    to use the following samplers:
    \begin{itemize}
    \item \texttt{base::Slice};
    \item \texttt{bugs::Dirichlet}.
    \end{itemize}
\end{enumerate}

We selected this restricted set of samplers to try and force the
\emph{jags-full-restricted} model to use the same set of samplers as the
\emph{jags-marg-inbuilt} and \emph{jags-marg} models. We did this to assess the
impact of marginalisation itself, rather than also the use of different
samplers (which JAGS chooses automatically, based on the model it is using).
The set of samplers was selected by turning all samplers `off' except for a
candidate set and looking for error messages when running the
\emph{jags-marg-inbuilt} and \emph{jags-marg} models.

\subsubsection{Dawid--Skene model with JAGS}

We have the \emph{jags-marg} model which is marginalised manually and the
unmarginalised \emph{jags-full} model which is allowed to use any samplers from
the \texttt{base} and \texttt{bugs} modules in JAGS. The
\emph{jags-full-restricted} model is unmarginalised but only allowed to use the
same set of samplers as for the \emph{jags-full-restricted} for Gaussian
mixture models. We selected this restricted set of samplers by going through
the same procedure for the Dawid--Skene model as we did for the Gaussian
mixture model (see above).

\subsection{Evaluating computational efficiency} \label{evaluate_efficiency}

We evaluated computational efficiency by looking at four quantities of
interest:

\begin{enumerate}
\item Computation time
\item Minimum effective sample size for the continuous parameters
\item Time per minimum effective sample \\
      (Computation time divided by Minimum effective sample size of continuous
      parameters)
\item The $\hat{R}$ statistic \citep{gelman_rubin}
\end{enumerate}
The way we measure these quantities is specified in \autoref{table:measure}.

\begin{table}
\begin{center}
\caption{Methods used for measuring the quantities to assess computational
         efficiency (see \autoref{evaluate_efficiency}).}
\label{table:measure}
\smallskip
\begin{tabular}{ p{3.5cm} p{6.5cm} p{6cm} }
\toprule
\textbf{Quantity} & \textbf{JAGS} & \textbf{Stan} \\
\midrule
\emph{Computation time} &
Using the \texttt{system.time()} function to measure the time elapsed running
\texttt{jags.model()} and \texttt{coda.samples()}. The quantity is the sum of
the two computation times. &
Using the \texttt{system.time()} function to measure the time elapsed running
\texttt{stan()}. \\
\addlinespace[1.2em]
\emph{Minimum effective sample size for the continuous parameters} &
\multicolumn{2}{p{12.5cm}}{(For both JAGS and Stan) Looking at
\texttt{ess\_bulk} and \texttt{ess\_tail} for all continuous parameters of
\texttt{posterior}'s \texttt{draws} object and taking the minimum. The numbers
were similar, so we only present the \texttt{ess\_bulk} measure in the result.}
\\
\addlinespace[1.2em]
\emph{Time per minimum effective sample} &
\multicolumn{2}{p{12.5cm}}{(For both JAGS and Stan) Computation time divided by
Minimum effective sample size of continuous parameters.} \\
\addlinespace[1.2em]
$\hat{R}$ &
\multicolumn{2}{p{12.5cm}}{(For both JAGS and Stan) Looking at \texttt{rhat}
for all \emph{continuous} parameters of \texttt{posterior}'s \texttt{draws}
object and taking the maximum.} \\
\bottomrule
\end{tabular}
\end{center}
\end{table}

We ran the models on each of the $5$ replicate datasets to obtain $5$
observations of each of the quantities above for each model--data pair. The
analysis was done with $3$ chains, a total of $3000$ iterations, with $1500$
iterations of warm up.


\section{Results}

\subsection{Gaussian mixture models}

\begin{figure}
\centering
\includegraphics[width=1\textwidth]{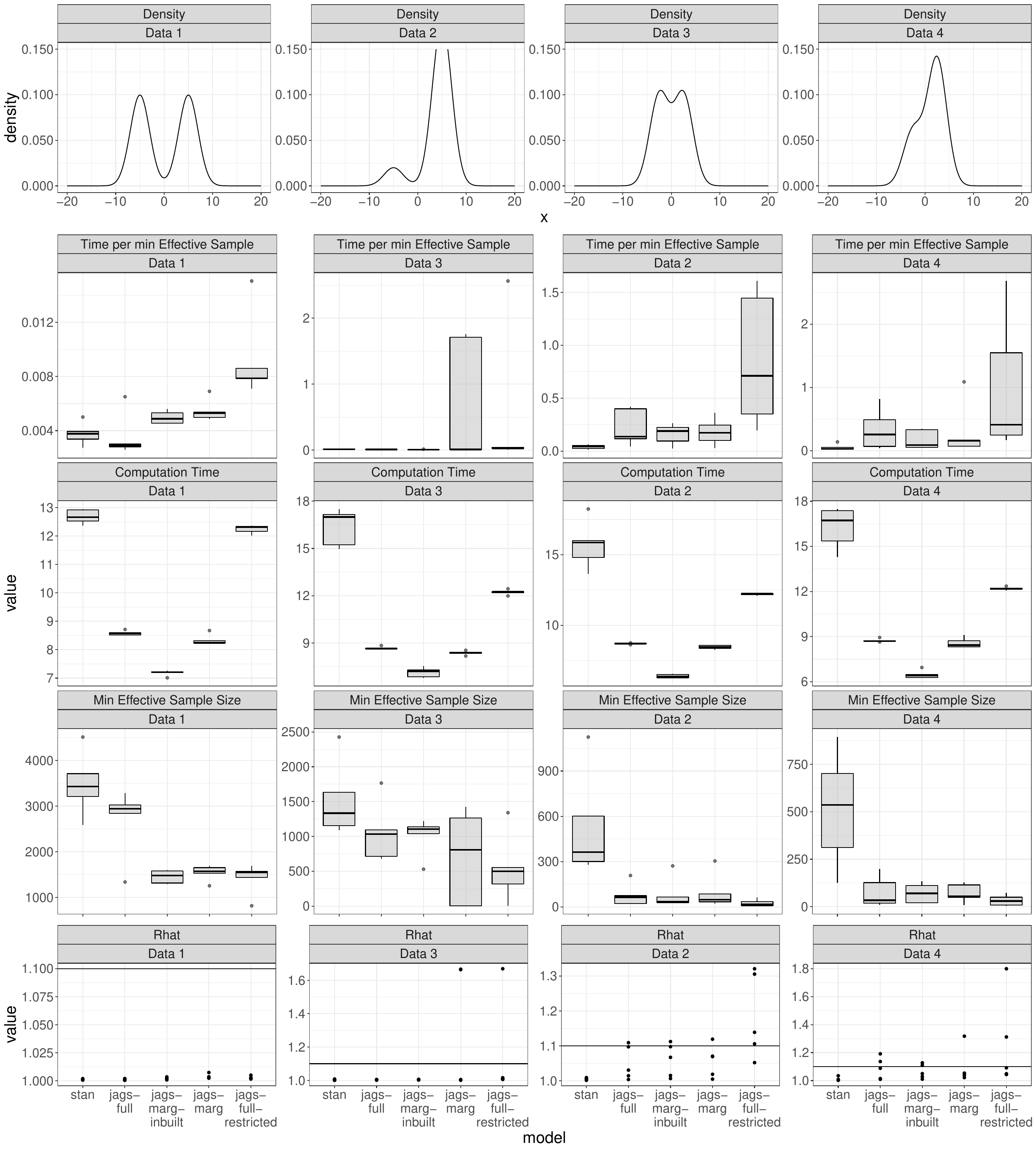}
\caption{Time per minimum effective sample, computation time, minimum effective
sample size and $\hat{R}$ when performing Bayesian inference of two-component
Gaussian mixture models. The analysis is done with $3$ chains, a total of
$3000$ iterations with $1500$ iterations of warm up. Each box in the boxplot
represents $5$ results. The $5$ results are obtained from $5$ trials of
analysis for each model--data pair. For each trial, we simulate one replicate
of the dataset and run analysis for \emph{stan}, \emph{jags-full},
\emph{jags-marg}, \emph{jags-marg-inbuilt} and \emph{jags-full-restricted}
respectively. The horizontal line in the $\hat{R}$ plots is the conventional
$1.1$ threshold.}
\label{fig:two-component-result}
\end{figure}

\begin{figure}
\centering
\includegraphics[width=1\textwidth]{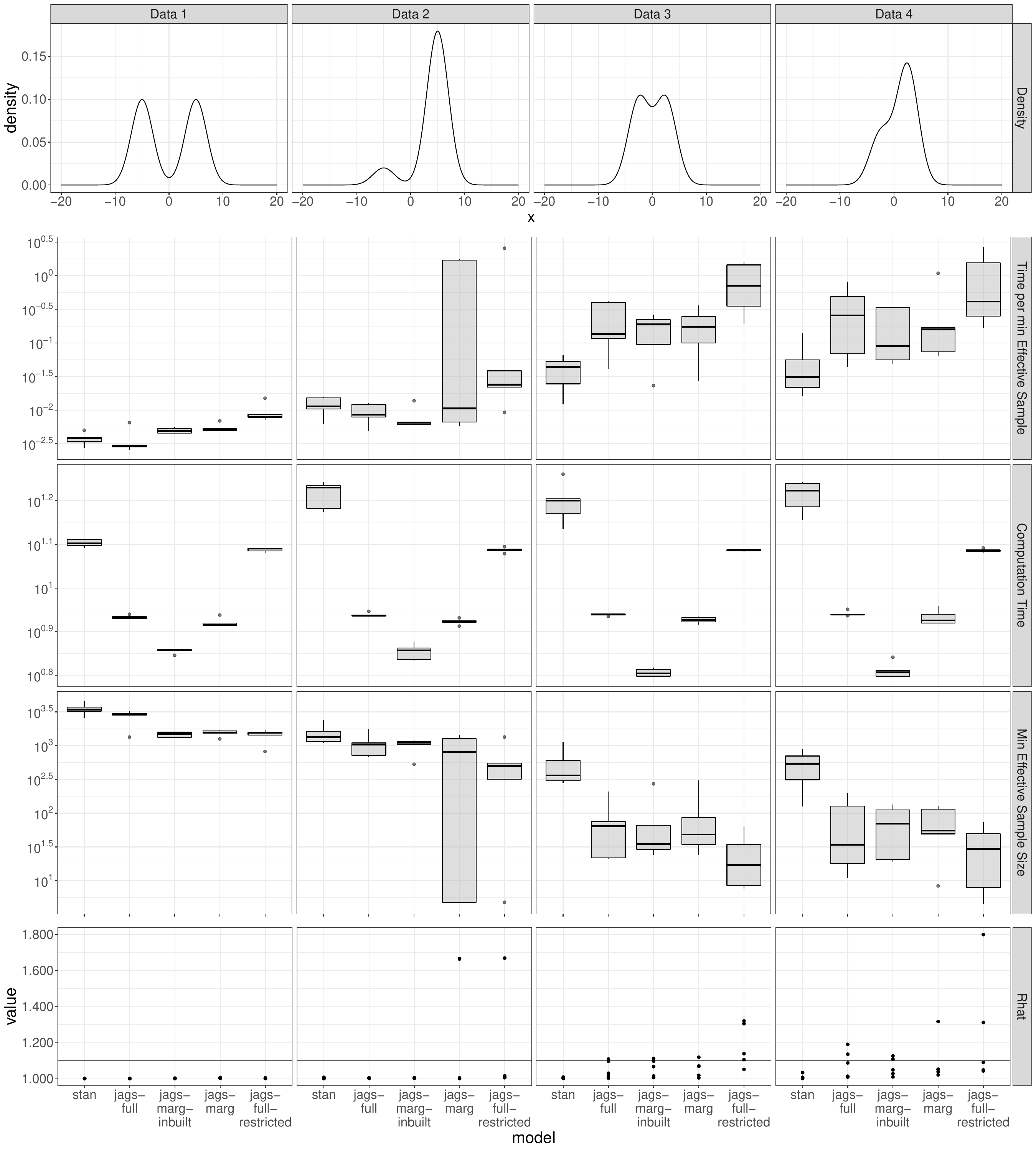}
\caption{Same as \autoref{fig:two-component-result} but with results in each
row shown on a common $\log_{10}$ scale to more easily compare across different
datasets.}
\label{fig:two-component-log-result}
\end{figure}

We used two plots to summarise results for two- and three-component Gaussian
mixture models respectively. \autoref{fig:two-component-result} and
\autoref{fig:three-component-result} shows results on the natural scale, while
\autoref{fig:two-component-log-result} and
\autoref{fig:three-component-log-result} shows results on a logarithmic scale
(base 10) with common limits on the vertical axes.
\autoref{fig:two-component-result} and \autoref{fig:three-component-result}
highlight differences between models with the same dataset;
\autoref{fig:two-component-log-result} and
\autoref{fig:three-component-log-result} demonstrate the differences across
different datasets.

The time per minimum effective sample measures the overall efficiency of each
method. It is clear from \autoref{fig:two-component-result} and
\autoref{fig:three-component-result} that \emph{stan} usually outperformed all
other models. Of the unrestricted JAGS models (i.e.\ all except
\emph{jags-full-restricted}), \emph{jags-full} tended to outperform slightly
when the (maximal) distance between components was large. When the (maximal)
distance between components was small, the difference between \emph{jags-full},
\emph{jags-marg-inbuilt} and \emph{jags-marg} was minor. The restricted model,
\emph{jags-full-restricted}, was consistently the worst performer across all
scenarios.

Considering computation time and minimum effective sample size separately,
although \emph{stan}'s computation time was often higher than (in the
two-component case) or comparable (in the three-component case) to those of the
JAGS models, it had the highest minimum effective sample size in all cases,
typically allowing it to be the most efficient overall. When the (maximal)
distance between components was small, \emph{jags-marg-inbuilt} and
\emph{jags-marg}, \emph{jags-full} were significantly faster than \emph{stan}
when it comes to computation time (but they typically also had smaller
effective sample sizes, thus did not typically outperform \emph{stan} in
overall efficiency).

The realised values of $\hat{R}$ demonstrate that \emph{stan} converges the
most consistently across all scenarios, whereas the JAGS models tend to perform
inconsistently on many of the datasets.

The results in \autoref{fig:two-component-log-result} and
\autoref{fig:three-component-log-result} suggest that is it `easier' (in the
sense that the time per minimum effective sample is lower) to draw posterior
samples when using Data~1 and Data~2 for both the two- and three-component
Gaussian mixture models than for the other data structures. What distinguishes
these datasets from the others is that their mixture components are distinct
and well-separated (they have the largest distance between components, and in
the three-component case the components are equidistant).

A highly imbalanced mixture proportion didn't have any substantial impact on
the relative performance of the models. This is evident from comparing Data $1$
and Data $2$ in \autoref{fig:two-component-log-result}. Although Data $2$ was
`harder' (in the sense that the time per minimum effective sample is higher)
for all models, the relative difference between models, in time per minimum
effective sample, is similar between the two datasets.

\begin{landscape}
\begin{figure}
\centering
\includegraphics[width=1.4\textwidth]{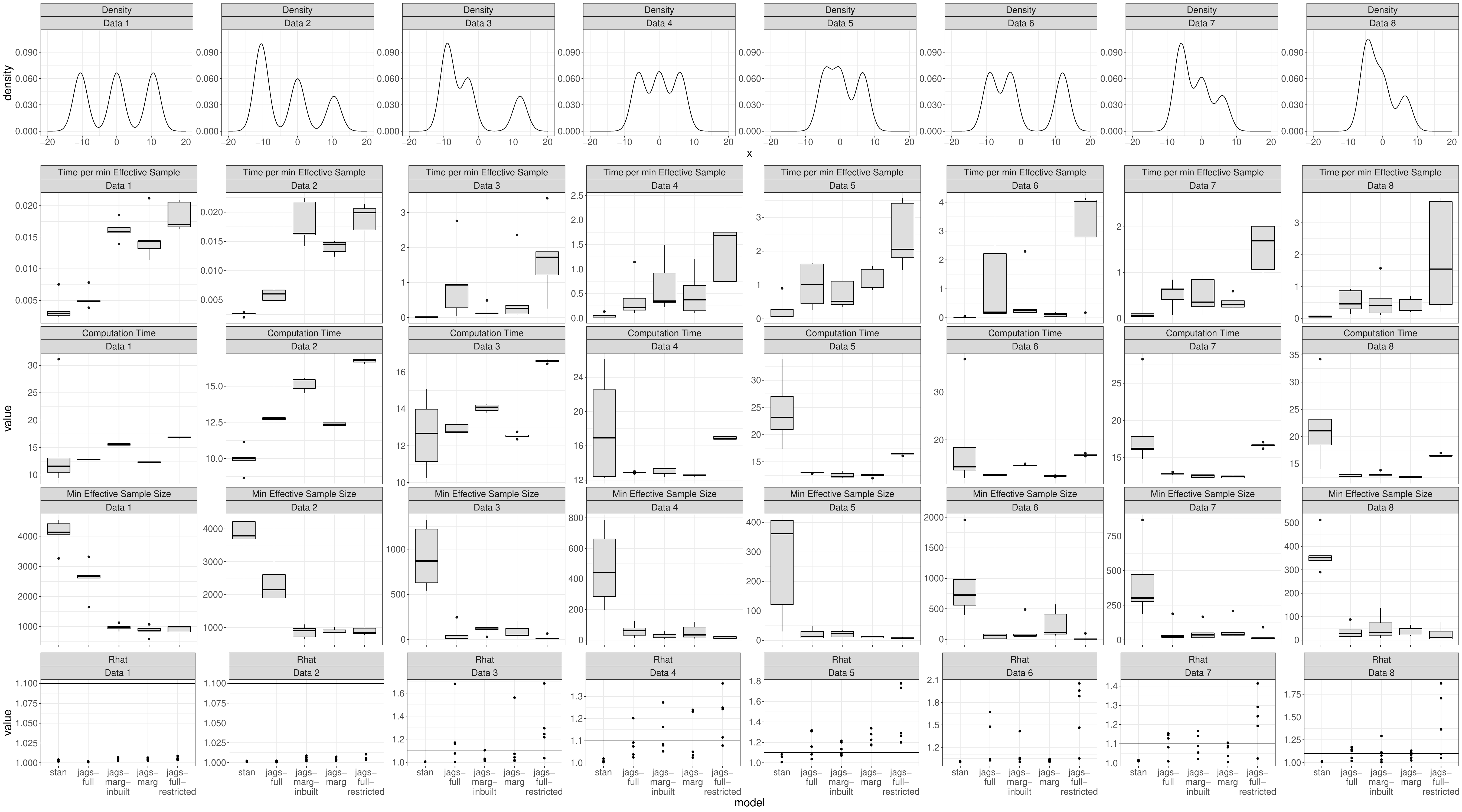}
\caption{Same as \autoref{fig:two-component-result} but now showing the results
for the three-component Gaussian mixture model.}
\label{fig:three-component-result}
\end{figure}

\begin{figure}
\centering
\includegraphics[width=1.4\textwidth]{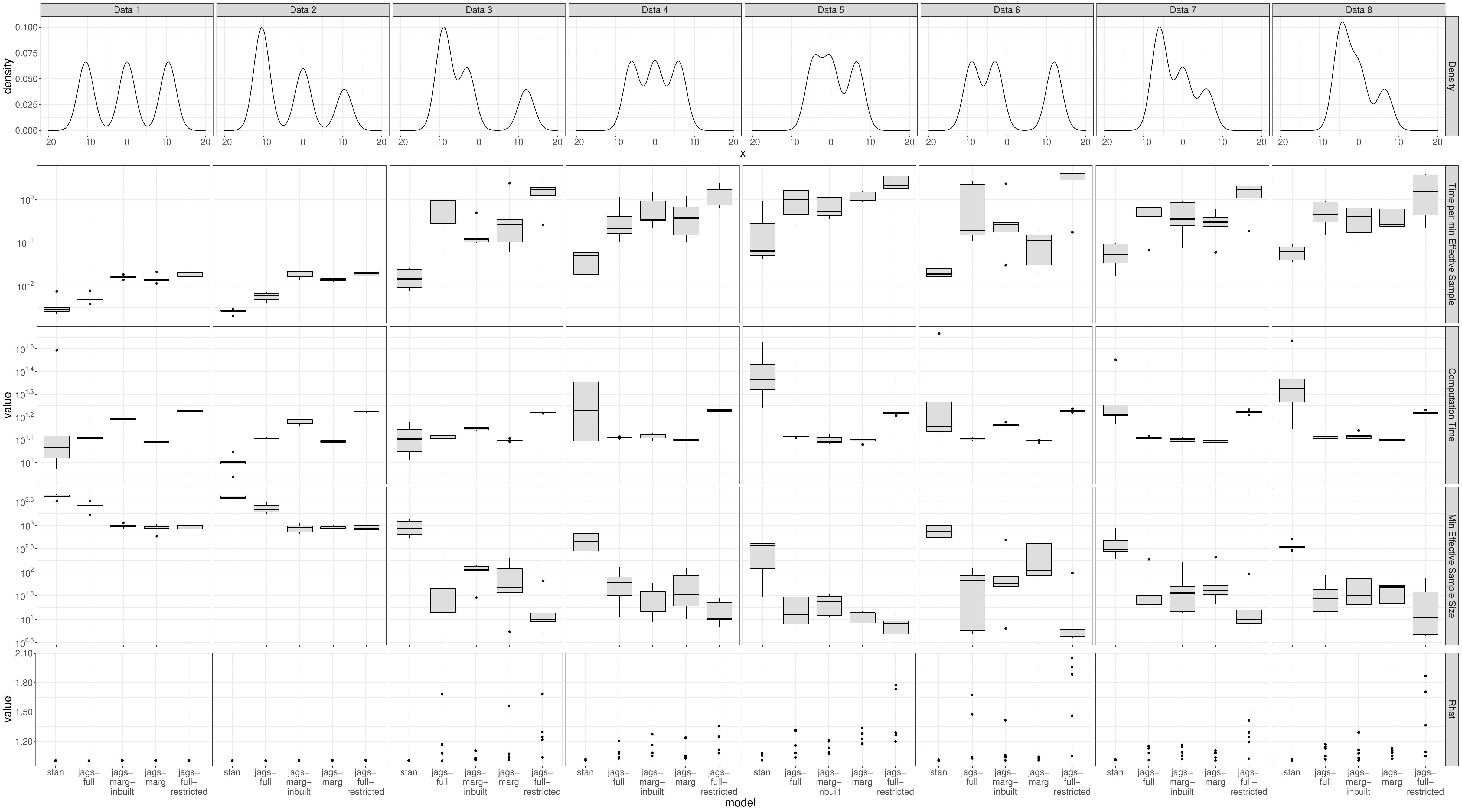}
\caption{Same as \autoref{fig:three-component-result} but with results in each
row shown on a common $\log_{10}$ scale to more easily compare across different
datasets.}
\label{fig:three-component-log-result}
\end{figure}
\end{landscape}

\subsection{Dawid--Skene model}

\begin{figure}
\centering
\includegraphics[width=0.7\textwidth]{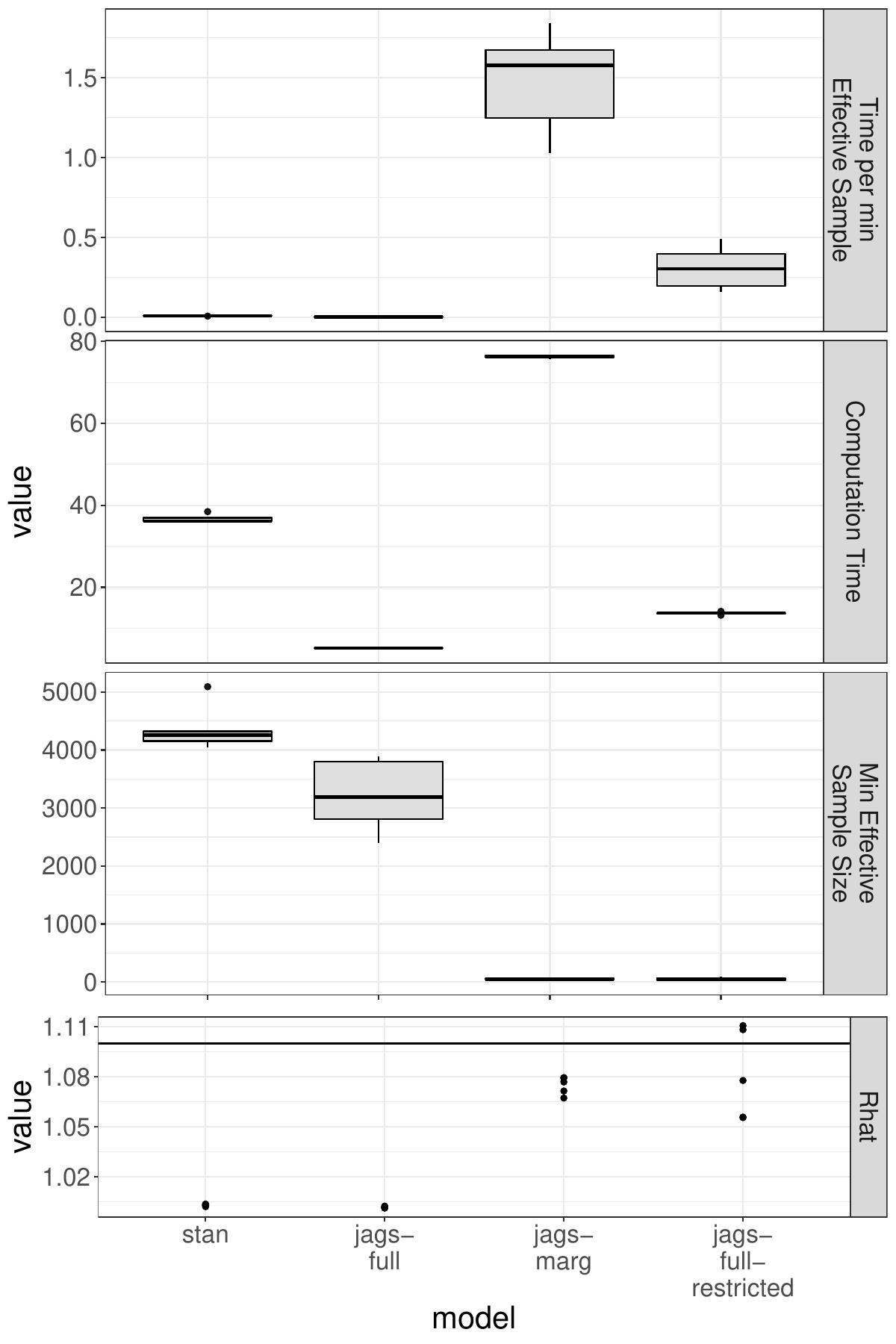}
\caption{Time per minimum effective sample, computation time, minimum effective
sample and $\hat{R}$ when performing Bayesian analysis on Dawid--Skene models
for categorical ratings. Each box in the boxplot represents $5$ results. The
horizontal line in the $\hat{R}$ plots is the $1.1$ threshold.}
\label{fig:Dawid-Skene-result}
\end{figure}

Unlike the Gaussian mixture models, \emph{jags-full} slightly outperformed
\emph{stan} as seen in the first row of \autoref{fig:Dawid-Skene-result}. The
\emph{stan} model had both a longer computation time and a larger minimum
effective sample size compared to \emph{jags-full}. However, the latter was so
fast that its time per minimum effective sample size was smaller.  The
$\hat{R}$ values demonstrate that \emph{stan} and \emph{jags-full} converged
the most consistently out of all models. The marginalised model,
\emph{jags-marg}, takes significantly longer computation time than all other
models which makes it the worst performer when looking at time per minimum
effective sample.

\subsection{Main conclusions}

Comparing the JAGS Gaussian mixture models, it is clear that marginalisation on
its own does not necessarily boosts computational efficiency.  The full model,
\emph{jags-full}, outperforms the two marginalised models
(\emph{jags-marg-inbuilt} and \emph{jags-marg}) in almost all cases in terms of
overall efficiency.

Furthermore, the software implementations and choice of samplers had a much
greater impact on computational efficiency.  We can see this in two ways.
First, by comparing the performance of \emph{jags-full} and
\emph{jags-full-restricted}.  The \emph{jags-full-restricted} model
consistently performed worse, and \emph{jags-full} was the most efficient out
of all JAGS models. This suggests that the boost from marginalisation is
insufficient to overtake the boost from using a more efficient sampler.
Second, we can compare the performance of \emph{stan} against that of the
marginalised JAGS models.  The \emph{stan} model (which is marginalised) was
more efficient than the JAGS models, highlighting the difference due to choice
of sampler and software implementation.

The results for Dawid--Skene suggest that marginalisation does not improve
computational efficiency at all. Through comparing the performance of
\emph{jags-marg} and \emph{jags-full-restricted}, the two models which we
believe are using the same set of samplers, we see that the unmarginalised
\emph{jags-full-restricted} model performs better than the \emph{jags-marg}
model in terms of time per minimum effective sample. This means the
unmarginalised model is in fact more computationally efficient even without the
boost from the software implementation. In contrast, the fact that the
marginalised \emph{stan} model had computational efficiency similar to
\emph{jags-full} shows that, once again, the sampler and software
implementation details are a more substantial factor in performance; in this
case, the benefit was enough to counteract the seeming disadvantage of
marginalisation.


\section{Discussion}

State-of-the-art software implementations of Markov chain Monte Carlo methods
for Bayesian computation and optimisation currently use sampling techniques
that require discrete parameters to be marginalised out of the full probability
model whose joint posterior is the target distribution. Sampling from the model
object that remains after marginalisation should be quicker and the process
should reach convergence sooner, but we know of only one study that
investigated this claim empirically. This question is important because
overcoming the computational burden when using iterative simulation is a
perennial challenge. If marginalisation of discrete parameters improves
efficiency then it could be used more generally in models for which it is not
strictly required. Marginalisation might then be viewed as one potential tool
in the implementation of any data analysis. The posterior expectation of the
marginalised discrete variable is often a quantity of interest in its own
right, and many of the properties of the distribution of the marginalised
parameter(s) can be identified and estimated from the sampled chains of values
for the remaining model parameters.

We presented numerical results reflecting the impact of marginalisation on the
computational efficiency for two- and three-component Gaussian mixture models
and the Dawid--Skene model for categorical ratings.  This investigation was
robust to the extent that we explored two software implementations of Markov
chain Monte Carlo techniques (JAGS and Stan) and directly compared marginalised
and non-marginalised models while holding constant other aspects of the
sampling procedure. Our results did not show that marginalisation on its own is
sufficient to boost performance.  Nevertheless, the most computationally
efficient implementation of our models was usually Stan, which requires
marginalisation.

One limitation of our study is that the models considered are either simple
(normal mixtures) or of only mid-level complexity (Dawid--Skene). The models
are much more analytically tractable and therefore better understood than the
Cormack--Jolly--Seber (CJS) model. This may be why an investigation of the CJS
model showed that code for marginalised models was anywhere from five to more
than 1,000 times faster than representations of the model that retained the
discrete parameters while maintaining essentially identical inferences
\citep{yackulic_need_2020}. The authors noted that ``understanding how
marginalisation works shrinks the divide between Bayesian and maximum
likelihood approaches to population models [and] allows users to minimise the
speed that is sacrificed when switching from a maximum likelihood approach''.
That is, a wider appreciation of the benefits of marginalisation has the
potential to promote the use of Bayesian statistical methods.

Another specific feature of the two- and three-component Gaussian mixture
models is that we used only informal experiments to conclude that using $10^2$
as the variance of the prior distribution for the $\mu$'s would boost
convergence performance. A more thorough investigation could and should be done
to choose a value for the variance that provides the greatest improvement in
convergence. We agree with \citet{yackulic_need_2020} that ``widespread
application of marginalisation in Bayesian population models will facilitate
more thorough simulation studies, comparisons of alternative model structures,
and faster learning''.



\addcontentsline{toc}{section}{References}
\bibliography{references}


\end{document}